\input{aipcheck}

\documentclass[
    ,final            
  ]
  {aipproc}

\layoutstyle{8x11single}

\begin{document}

\title 
 [Perfect Abelian dominance of confinement in SU(3) lattice QCD]
 {Perfect Abelian dominance of confinement in 
  quark-antiquark potential in SU(3) lattice QCD}

\classification{12.38.Aw, 12.38.Gc}
\keywords{confinement, lattice QCD, quark potential, dual superconductor}

\author{Hideo Suganuma}{
  address={Department of Physics, Kyoto University, 
  Kitashirakawaoiwake, Sakyo, Kyoto 606-8502, Japan},
  email={suganuma@scphys.kyoto-u.ac.jp},
}

\iftrue
\author{Naoyuki Sakumichi}{
  address={Theoretical Research Division, Nishina Center, 
  RIKEN, Wako, Saitama 351-0198, Japan},
  email={naoyuki.sakumichi@riken.jp},
}

\fi

\copyrightyear  {2014}

\begin{abstract}
In the context of the dual superconductor picture 
for the confinement mechanism, 
we study maximally Abelian (MA) projection of quark confinement in SU(3) 
quenched lattice QCD with $32^4$ at $\beta$=6.4 (i.e., $a \simeq$ 0.058 fm). 
We investigate the static quark-antiquark potential $V(r)$, 
its Abelian part $V_{\rm Abel}(r)$ and its off-diagonal part 
$V_{\rm off}(r)$, respectively, from the on-axis lattice data. 
As a remarkable fact, we find almost perfect Abelian dominance 
for quark confinement, i.e., 
$\sigma_{\rm Abel} \simeq \sigma$ for the string tension, 
on the fine and large-volume lattice.
We find also a nontrivial summation relation of 
$V(r) \simeq V_{\rm Abel}(r) + V_{\rm off}(r)$. 
\end{abstract}

\date{\today}

\maketitle

\section{Introduction}

In 1966, Yoichiro Nambu \cite{N66} first proposed the SU(3) gauge theory, 
i.e., quantum chromodynamics (QCD), 
as a candidate for the fundamental theory of the strong interaction, 
just after the introduction of the color quantum number \cite{HN65}.
In 1973, the asymptotic freedom of QCD was theoretically proven \cite{GWP73}, 
and QCD was established as the fundamental theory of the strong interaction 
via the great success of perturbative QCD to high-energy hadron reactions.
Even at present, however, analytical treatment of QCD 
is quite difficult in the infrared region, 
because of its strong-coupling nature there. 
Actually, in spite of its simple form, QCD creates thousands of hadrons 
and leads to various interesting nonperturbative phenomena 
such as color confinement \cite{N74} and 
dynamical chiral-symmetry breaking \cite{NJL61}.
Since the first application \cite{C7980} of lattice QCD Monte Carlo 
simulations in 1979, lattice QCD has been applied as 
the direct numerical analysis for nonperturbative QCD \cite{R12}.

Among the nonperturbative properties of QCD, color confinement is 
one of the most important subjects remaining in elementary particle physics, 
and is also an extremely difficult mathematical problem.
The difficulty is considered to originate from non-Abelian dynamics 
and nonperturbative features of QCD, which are largely different from QED. 
However, it is not clear whether quark confinement is peculiar to 
the non-Abelian nature of QCD or not.

In 1970's, Nambu, 't~Hooft, and Mandelstam proposed 
an interesting idea that quark confinement might be physically 
interpreted using the dual version of the superconductivity \cite{N74}. 
In the ordinary superconductor, 
Cooper-pair condensation leads to the Meissner effect, 
and the magnetic flux is excluded or squeezed like a 
quasi-one-dimensional tube as the Abrikosov vortex, 
where the magnetic flux is quantized topologically. 
On the other hand, from the Regge trajectory of mesons and baryons, 
quark confining force in hadrons 
is characterized by a universal physical quantity 
of the string tension $\sigma \simeq$ 0.89 GeV/fm \cite{N6970}, 
and lattice QCD calculations also indicates one-dimensional squeezing 
of the color-electric flux in the QCD vacuum \cite{R12}. 
Therefore, the QCD vacuum could be regarded as the dual version 
of the superconductor based on above similarities 
on the low-dimensionalization of the quantized flux between charges. 

In the dual-superconductor picture for the QCD vacuum, 
the squeezing of the color-electric flux between quarks 
is realized by the dual Meissner effect, 
as the result of condensation of color-magnetic monopoles, 
which is the dual version of the electric charge as the Cooper pair.
However, there are two large gaps between QCD and the 
dual-superconductor picture \cite{IS9900}.

\begin{enumerate}
\item
The dual-superconductor picture is based on the Abelian gauge theory 
subject to the Maxwell-type equations, where electro-magnetic duality is 
manifest, while QCD is a non-Abelian gauge theory.
\item
The dual-superconductor picture requires condensation of 
color-magnetic monopoles as the key concept, while QCD does not 
have such a monopole as the elementary degrees of freedom.
\end{enumerate}
On the connection between QCD and the dual-superconductor scenario, 
't~Hooft proposed the concept of the Abelian gauge fixing \cite{tH81},
a partial gauge fixing which only remains 
Abelian gauge degrees of freedom in QCD.
By the Abelian gauge fixing, QCD reduces into an Abelian gauge theory, 
where off-diagonal gluons behave as charged matter fields 
similar to $W^\pm_\mu$ in the Weinberg-Salam model 
and give color-electric current $j_\mu$ 
in terms of the residual Abelian gauge symmetry.
Remarkably in the Abelian gauge, 
color-magnetic monopoles appear as topological objects 
corresponding to the nontrivial homotopy group 
$\Pi_2({\rm SU}(N_c)/U(1)^{N_c-1})={\bf Z}^{N_c-1}_\infty$ 
in a similar manner to the GUT monopole \cite{tH81,SST95}.
In general, the monopole appears as a topological defect or 
a singularity in a constrained Abelian gauge manifold embedded 
in the compact (and at most semi-simple) non-Abelian gauge manifold.

In fact, by the Abelian gauge fixing, QCD reduces into an Abelian 
gauge theory including both the electric current $j_\mu$ and 
the magnetic current $k_\mu$, which is expected to give a theoretical basis 
of the dual-superconductor scheme for the confinement mechanism 
\cite{SST95,EI82,S88,K98}.
In particular, in the maximally Abelian (MA) gauge \cite{KSW87}, 
which is a special Abelian gauge, the off-diagonal gluon has 
a large effective mass of about 1GeV \cite{AS99,GIS12}, 
and Abelian dominance of quark confinement is approximately 
observed in lattice QCD \cite{SY90,BBMS96,P97}.
Then, in the MA gauge, QCD becomes an infrared Abelian gauge theory 
including the magnetic monopole current $k_\mu$ \cite{KSW87} 
together with the electric current $j_\mu$. 
By the Hodge decomposition, the QCD system can be divided into 
the monopole part ($k_\mu \ne 0$, $j_\mu=0$) and 
the photon part ($j_\mu \ne 0$, $k_\mu = 0$). 
The lattice QCD studies have shown that 
the monopole part has quark confinement \cite{SNW94},
chiral symmetry breaking \cite{M95W95} and instantons \cite{STSM95},
while the photon part does not have all of them. (See Fig.1.)
Monopole condensation is also suggested by 
long entangled monopole worldlines \cite{KSW87,SNW94} 
and the magnetic screening \cite{SATI00}.

Thus, in the MA gauge, the infrared QCD system resembles 
a dual superconductor, and the color-magnetic monopole seems to 
carry the essence of nonperturbative QCD.
However, such lattice studies were mainly performed in simplified 
SU(2) color QCD 
\cite{IS9900,KSW87,AS99,SY90,BBMS96,P97,SNW94,M95W95,STSM95,SATI00}, 
and there are only a few 
pioneering studies \cite{STW02,DIK04,L04} on the Abelian dominance of 
quark confinement in SU(3) color QCD in the real world.

\begin{figure}[h]
\centering
\includegraphics[width=15cm,clip]{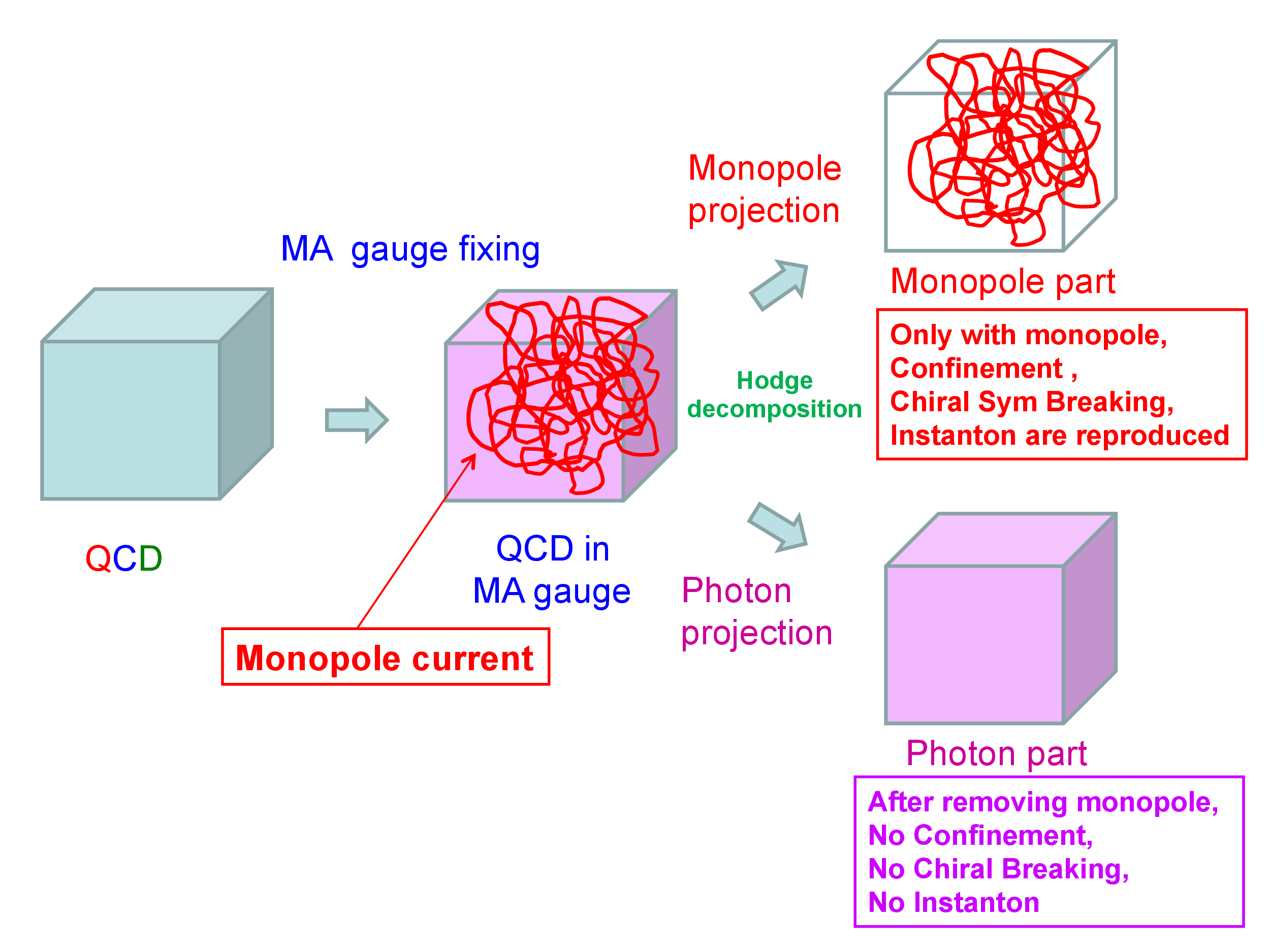}
\caption{
A schematic figure of the dual superconductor picture from QCD 
in the MA gauge. 
In the MA gauge, QCD becomes Abelian-like because of a 
large off-diagonal gluon mass of about 1GeV \cite{AS99,GIS12}, and 
monopole currents topologically appear \cite{KSW87}. 
By the Hodge decomposition, the QCD system in the MA gauge 
can be divided into the monopole part ($k_\mu \ne 0$, $j_\mu=0$) 
and the photon part ($j_\mu \ne 0$, $k_\mu=0$). 
The monopole part has quark confinement \cite{SNW94},
chiral symmetry breaking \cite{M95W95} and instantons \cite{STSM95},
while the photon part does not have all of them, 
as shown in lattice QCD. 
}
\end{figure}

In this paper, we study Abelian dominance of quark confinement 
in the MA gauge in SU(3) color QCD at the quenched level 
on a fine and large-volume lattice of $32^4$ at $\beta=6.4$.
After the calculation of the quark-antiquark (Q$\bar{\rm Q}$) potential $V(r)$ and 
its Abelian part $V_{\rm Abel}(r)$ in the MA gauge, 
we consider the fit analysis for them and compare their infrared behavior, 
and thus investigate quantitatively Abelian dominance 
of quark confinement in SU(3) QCD.

\section{Formalism}

In this section, we briefly review the maximally Abelian (MA) gauge, 
the Cartan decomposition, and MA projection 
in ${\rm SU}(N_c)$ QCD in both continuum and lattice formalism.

\subsection{SU(N) continuum QCD}

To begin with, we consider Euclidean continuum QCD 
and the gluon field $A_\mu (x) \in {\rm su}(N_c)$.
The MA gauge is a special Abelian gauge 
which minimizes the ``global off-diagonal gluon amplitude''
\cite{IS9900},
\begin{eqnarray}
R_{\rm off} [A_\mu ( \cdot )] \equiv \int d^4x \ {\rm tr}
[\hat D_\mu ,\vec H][\hat D^\mu ,\vec H]^\dagger
={g^2 \over 2} \int d^4x \sum_\alpha |A_\mu ^\alpha (x)|^2,
\end{eqnarray}
with the ${\rm SU}(N_c)$ covariant derivative 
$\hat D_\mu \equiv \hat \partial_\mu+igA_\mu $ and 
the Cartan decomposition 
$A_\mu (x)=\vec A_\mu (x) \cdot \vec H +\sum_\alpha A_\mu^\alpha (x)E^\alpha $.
Thus, in the MA gauge, the off-diagonal gluon amplitude 
$|A_\mu ^\alpha (x)|$ is forced to be globally minimized 
by the gauge transformation, 
and hence the gluon field $A_\mu (x) \equiv A_\mu ^a(x)T^a$ 
maximally approaches the Abelian gauge field 
$\vec A_\mu (x) \cdot \vec H$. 

Since $R_{\rm off}$ is gauge-transformed by $\Omega \in G$ as 
\begin{eqnarray}
R_{\rm off} \rightarrow 
R_{\rm off}^\Omega 
=\int d^4x \ {\rm tr}
[\Omega \hat D_\mu \Omega ^\dagger, \vec H]
[\Omega \hat D^\mu \Omega ^\dagger, \vec H]^\dagger 
=\int d^4x \ {\rm tr}
[\hat D_\mu ,\Omega ^\dagger \vec H \Omega ]
[\hat D^\mu ,\Omega ^\dagger \vec H \Omega ]^\dagger,
\end{eqnarray}
the MA gauge fixing condition is obtained as \cite{IS9900}
\begin{eqnarray}
[\vec H, [\hat D_\mu , [\hat D^\mu , \vec H]]]=0
\end{eqnarray}
from the infinitesimal gauge transformation of $\Omega $.

In the MA gauge, the gauge group 
$G \equiv {\rm SU}(N_c)_{\rm local}$ 
is reduced into ${\rm U(1)}_{\rm local}^{N_c-1} 
\times {\rm Weyl}_{\rm global}$,
where the global Weyl symmetry is the subgroup of ${\rm SU}(N_c)$ 
relating the permutation of the basis in the fundamental 
representation \cite{IS9900}. 
In continuum QCD, MA projection is a simple replacement of 
the non-Abelian gluon field 
$A_\mu (x) \equiv A_\mu ^a(x)T^a \in {\rm su}(N_c)$ 
by its Abelian part $\vec A_\mu (x) \cdot \vec H$. 

\subsection{SU(N) lattice QCD}

Next, let us consider Euclidean lattice QCD formalism, where 
the gauge field is described by the link-variable 
$U_\mu(s) =e^{iagA_\mu(s)} \in {\rm SU}(N_c)$, 
with the lattice spacing $a$ and the gauge coupling $g$.
In lattice QCD, the MA gauge is defined by maximizing 
the diagonal element of the link-variable, 
\begin{equation}
R_{\rm MA}[U_\mu(\cdot)]\equiv \sum_{s} \sum_{\mu=1}^4 
{\rm tr}\left( U_\mu^\dagger(s)\vec H U_\mu(s)\vec H\right), 
\label{eq:RMA}
\end{equation}
under the ${\rm SU}(N_c)$ gauge transformation 
\begin{equation}
U_\mu(s)\rightarrow U_\mu^\Omega(s)\equiv 
\Omega(s)U_\mu(s)\Omega^\dagger(s+\hat\mu), 
\end{equation}
with the gauge function $\Omega(s) \in {\rm SU}(N_c)$.

Corresponding to the Cartan decomposition, 
the ${\rm SU}(N_c)$ link-variable 
$U_\mu (s)$ is factorized 
into a maximal-torus-subgroup element 
$u_\mu(s) \in {\rm U(1)}^{N_c-1}$ 
and a coset-space element $M_\mu(s) \in {\rm SU}(N_c)/{\rm U(1)}^{N_c-1}$ as
\begin{eqnarray}
U_\mu (s)=M_\mu (s)u_\mu (s) \in G \equiv {\rm SU}(N_c), 
\end{eqnarray}
with 
\begin{eqnarray}
u_\mu (s) = e^{i \vec \theta _\mu (s) \cdot \vec H}
\in H \equiv {\rm U(1)}^{N_c-1}, \qquad
M_\mu (s) = e^{i\Sigma _\alpha \theta _\mu ^\alpha (s)E^\alpha} 
\in G/H.
\end{eqnarray}

In the MA gauge, there remains the residual 
${\rm U(1)}^{N_c-1}$ gauge symmetry, 
because $R_{\rm MA}$ in Eq.~(\ref{eq:RMA}) is invariant under 
the ${\rm U(1)}^{N_c-1}$ gauge transformation 
\begin{equation}
U_\mu(s)\rightarrow U_\mu^\omega(s)\equiv 
\omega(s)U_\mu(s)\omega^\dagger(s+\hat\mu)
\label{eq:ResGS}
\end{equation}
with $\omega(s) \in {\rm U(1)}^{N_c-1}$.
By the residual gauge transformation (\ref{eq:ResGS}), 
$u_\mu(s)$ and $M_\mu(s)$ transform as
\begin{eqnarray}
u_\mu(s)&\rightarrow& u_\mu^\omega(s)\equiv 
\omega(s)u_\mu(s)\omega^\dagger(s+\hat\mu), 
\label{eq:ResGT}
\\ 
M_\mu(s)&\rightarrow& M_\mu^\omega(s)\equiv 
\omega(s)M_\mu(s)\omega^\dagger(s),
\end{eqnarray}
where $M_\mu(s)$ keeps the form of 
$e^{i \sum_\alpha \theta_\mu^\alpha(s)E^\alpha}\in G/H$.
Thus, the Abelian link-variable 
$u_\mu (s) \in H={\rm U(1)}^{N_c-1}$ behaves as 
the gauge field in ${\rm U(1)}^{N_c-1}$ lattice gauge theory, 
which is similar to the compact QED, while 
the off-diagonal factor $M_\mu (s) \in G/H$ 
behaves as a charged matter field 
in terms of the residual Abelian gauge symmetry 
${\rm U(1)}^{N_c-1}_{\rm local}$.

In the lattice QCD, MA projection is defined by the replacement of 
\begin{eqnarray}
U_\mu (s) \in G \quad \rightarrow \quad u_\mu (s) \in H.
\end{eqnarray}

\section{Lattice setup and potential measurement}

Using the standard plaquette action, 
we perform SU(3) quenched lattice QCD Monte Carlo calculations 
on a fine and large-volume lattice of $32^4$ at $\beta \equiv 6/g^2 =6.4$, 
i.e., $a\simeq 0.058(7)$ fm. Here, the lattice spacing $a$ is determined 
so as to reproduce the string tension $\sigma = 0.89$ GeV/fm 
in the inter-quark potential. 
For comparison, we also investigate a smaller and coarser lattice 
of $16^3 \times 32$ at $\beta=6.0$, i.e., $a \simeq 0.105(14)$ fm.
After a thermalization of $20000$ sweeps, 
we sample the gauge configuration every $500$ sweep, and 
generate $70$ and $250$ gauge configurations 
for $\beta = 6.4$ and $6.0$, respectively.
On the error estimate, we use the jackknife method.

The (ground-state) quark-antiquark (Q$\bar{\rm Q}$) potential $V(r)$ 
at the inter-quark distance $r$ is calculated as 
\begin{eqnarray}
V(r) = - \lim_{T\rightarrow \infty}\frac{1}{T}\ln \left\langle 
W_C \left[ U_\mu(s) \right] \right\rangle
\end{eqnarray}
from the Wilson loop
\begin{equation}
W_C \left[ U_\mu(s) \right]
\equiv {\rm tr} \{ \prod_C U_\mu (s) \}.
\end{equation}
Here, $C$ denotes the $r \times T$ rectangular loop, 
and $\langle \cdots \rangle$ means the statistical average 
over the gauge configurations. 
For the accurate calculation of the Q$\bar{\rm Q}$ potential, 
we adopt the gauge-invariant smearing method, 
which reduces the excited-state components and enhances 
the ground-state overlap in the Q$\bar{\rm Q}$ system \cite{BSS93,TSNM02}.

The Abelian part $V_{\rm Abel} (r)$, i.e., 
MA projection of the Q$\bar{\rm Q}$ potential, is obtained as 
\begin{eqnarray}
V_{\rm Abel} (r) = - \lim_{T\rightarrow \infty}\frac{1}{T}\ln \left\langle 
W_C \left[ u_\mu(s) \right] \right\rangle
\end{eqnarray}
from the Abelian Wilson loop $W_C \left[ u_\mu(s)\right]$ in the MA gauge, 
which is invariant under 
the residual Abelian gauge transformation~(\ref{eq:ResGT}) and 
the global Weyl transformation \cite{IS9900}.

Also, we define the off-diagonal part 
$V_{\rm off}(r)$ of the Q$\bar{\rm Q}$ potential by 
\begin{eqnarray}
V_{\rm off} (r) = - \lim_{T\rightarrow \infty}\frac{1}{T}\ln \left\langle 
W_C \left[ M_\mu(s) \right] \right\rangle 
\end{eqnarray}
from the off-diagonal Wilson loop $W_C \left[ M_\mu(s)\right]$ in the MA gauge 
with the ${\rm U(1)}^{N_c-1}$ Landau gauge, which maximizes 
\begin{equation}
R_L [u_\mu(\cdot)]\equiv \sum_{s} \sum_{\mu=1}^4 
{\rm Re} \, {\rm tr} \left( u_\mu(s) \right)
\end{equation}
under the residual gauge transformation~(\ref{eq:ResGT}).
(For the argument of perfect Abelian dominance, 
$V_{\rm off}(r)$ is not needed.)

For the Q$\bar{\rm Q}$ potential, 
we here investigate the on-axis data only, and 
consider the inter-quark distance $2 \leq r \leq 17$ for $\beta =6.4$, 
and $1 \leq r \leq 8$ for $\beta =6.0$ in the lattice unit.
(In Ref.\cite{SS14}, we investigate both on-axis and off-axis data 
from several different lattices with much more gauge configurations, 
and find that the main conclusion is almost the same.)

\section{SU(3) lattice QCD results}

We show in Fig.2(a) the SU(3) lattice QCD result of
the Q$\bar{\rm Q}$ potential $V(r)$, 
the Abelian part $V_{\rm Abel}(r)$ in the MA gauge, 
and the off-diagonal part $V_{\rm off}(r)$. 
The main panels of Fig.2 show the on-axis (integer $r$) data 
on the fine and large-volume lattice of $32^4$ at $\beta =6.4$, 
and the insets those on 
a smaller and coarser lattice of $16^3 \times 32$ at $\beta =6.0$.

\vspace{0.4cm}

\begin{figure}[h]
\centering
\includegraphics[width=16.5cm,clip]{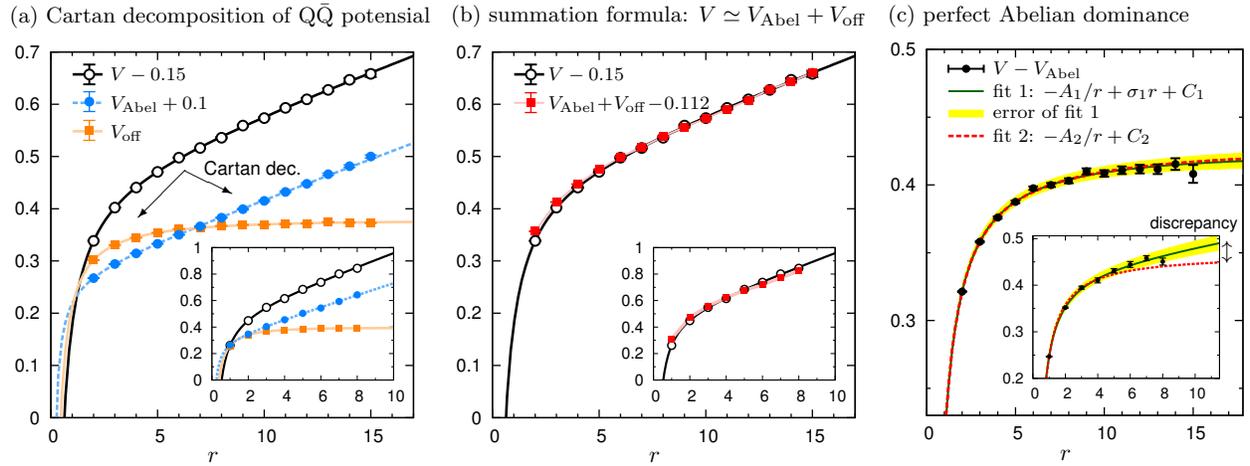}
\caption{
(a) The SU(3) lattice QCD result of
the Q$\bar{\rm Q}$ potential $V(r)$ (open circles), 
the Abelian part $V_{\rm Abel}(r)$ (filled circles)
in the MA gauge, 
and the off-diagonal part $V_{\rm off}(r)$ (squares). 
For each part, the best-fit curve with Eq.~(\ref{eq:fit}) is added.
(b) Comparison between $V_{\rm Abel}(r) + V_{\rm off}(r)$ (squares) and 
$V(r)$ (open circles), except for an irrelevant constant.
Their agreement indicates a summation formula of Eq.~(\ref{eq:sum}).
(c) Fit analysis of $V(r) - V_{\rm Abel}(r)$ (filled circles) 
to demonstrate perfect Abelian dominance of quark confinement.
The solid curve denotes the best fit with the Coulomb-plus-linear Ansatz 
of Eq.~(\ref{eq:fit}).
The dotted curve denotes the best fit with the pure Coulomb Ansatz. 
For (a), (b), and (c), the main panels show the on-axis (integer $r$) data 
on the fine and large-volume lattice of $32^4$ at $\beta =6.4$, 
and the insets those on 
a smaller and coarser lattice of $16^3 \times 32$ at $\beta =6.0$.
}
\end{figure}

\subsection{Fit analysis with Coulomb-plus-linear Ansatz}

It is known from lattice QCD studies \cite{R12,BSS93,TSNM02} that 
the Q$\bar{\rm Q}$ potential $V(r)$ is well reproduced 
by a sum of the Coulomb and the linear confinement terms,
\begin{equation}
V(r) = - \frac{A}{r}+\sigma r +C,
\label{eq:fit}
\end{equation}
with the string tension $\sigma$, 
the color-Coulomb coefficient $A$, and an irrelevant constant $C$.

Then, we carry out the fit analysis 
for $V(r)$, $V_{\rm Abel}(r)$, and $V_{\rm off}(r)$, respectively,
based on the Coulomb-plus-linear Ansatz of Eq.~(\ref{eq:fit}).
We summarize the best-fit parameter set $(\sigma, A, C)$ 
with $\chi^2/N_{\rm df}$ for the main lattice ($32^4$, $\beta =6.4$) and 
($16^3 \times 32$, $\beta =6.0$) 
in Tables~1 and 2, respectively.
In Fig.2(a), we have added the best-fit curves with Eq.~(\ref{eq:fit}).

From the fit analysis, we find for both lattices that 
the Abelian part $V_{\rm Abel}(r)$, i.e., 
MA-projected Q$\bar{\rm Q}$ potential, 
can be well fitted with the Coulomb-plus-linear Ansatz,
as well as $V(r)$.
We also find that the off-diagonal part $V_{\rm off}(r)$ 
has almost zero string tension, i.e., $\sigma_{\rm off} \simeq 0$, 
and is approximated as a pure Coulomb potential \cite{SS14}.

\vspace{0.4cm}

\begin{table}[h]
\caption{
Fit analysis with the Coulomb-plus-linear Ansatz 
for on-axis data of the Q$\bar{\rm Q}$ potentials, 
$V(r)$, $V_{\rm Abel}(r)$, and $V_{\rm off}(r)$,
on a fine and large-volume lattice of $32^4$ 
at $\beta =6.4$, i.e., $a\simeq 0.058(7)$~fm. 
For each potential, the best-fit parameter set $(\sigma, A, C)$ 
in the functional form of Eq.~(\ref{eq:fit}) 
and $\chi^2/N_{\rm df}$ are listed.
}
\begin{tabular}{ccccc}
\hline
 & $\sigma$ $ [a^{-2}]$ & $A$ & $C$ $ [a^{-1}]$ & $\chi^2/N_{\rm df}$ \\
\hline
$V$ \ & \,\, 0.01507(20) & 0.290(3) & 0.603(2) & 1.14 \\
$V_{\rm Abel}$  & \,\, 0.01528(14) & 0.067(2)  & 0.170(1) & 1.17 \\
$V_{\rm off}$ & $-$0.00038(11) & 0.179(3) & 0.392(1) & 2.41\\
$V_{\rm Abel} + V_{\rm off}$ & \,\, 0.01510(15) & 0.242(3) & 0.560(2) & 0.94\\
\hline
\end{tabular}
\end{table}

\begin{table}[h]
\caption{
Fit analysis with the Coulomb-plus-linear Ansatz 
for on-axis data of the Q$\bar{\rm Q}$ potentials 
 ($V$, $V_{\rm Abel}$, $V_{\rm off}$)
on a smaller and coarser lattice of $16^3 \times 32$ 
at $\beta =6.0$, i.e., $a\simeq 0.105(14)$ fm. 
}
\begin{tabular}{clccc}
\hline
 & \quad $\sigma$ $ [a^{-2}] $ & 
$A$ & $C$ $ [a^{-1}]$ & $\chi^2/N_{\rm df}$ \\
\hline
$V$            & \, 0.0499(9)      & 0.275(4) & 0.637(5) & 1.26 \\
$V_{\rm Abel}$ & \, 0.0443(5)      & 0.075(2) & 0.196(2) & 1.38 \\
$V_{\rm off}$  & $-$0.0019(3)      & 0.166(2) & 0.427(2) & 3.51\\
$V - V_{\rm Abel}$ & \, 0.0058(10) & 0.200(4) & 0.440(5) & 1.05\\
\hline
\end{tabular}
\end{table}

\subsection{Perfect Abelian dominance of quark confinement}

As a remarkable fact, we find almost perfect Abelian dominance 
\cite{SS14} of the string tension, $\sigma_{\rm Abel} \simeq \sigma$, 
on the fine and large-volume lattice of $32^4$ at $\beta =6.4$, 
as shown in Fig.2(a) and in Table~1.
On the other hand, as shown in Table~2, 
we find only approximate Abelian dominance, 
$\sigma_{\rm Abel} \simeq 0.9 \sigma$, 
on the smaller and coarser lattice of $16^3 \times 32$ at $\beta =6.0$, 
which is consistent with the previous studies \cite{STW02,DIK04}.

To demonstrate the perfect Abelian dominance clearly, we perform 
the two fit analyses for $V(r) - V_{\rm Abel}(r)$ on 
the $32^4$ lattice at $\beta =6.4$ 
in terms of i) Coulomb-plus-linear Ansatz of Eq.~(\ref{eq:fit}) and 
ii) pure Coulomb Ansatz of Eq.~(\ref{eq:fit}) with $\sigma =0$, 
respectively.
As shown in Table~3, the former fit indicates no difference 
between the string tensions in $V(r)$ and $V_{\rm Abel}(r)$ 
with almost perfect precision, and the latter fit indicates that 
$V(r) - V_{\rm Abel}(r)$ is well described 
by a pure Coulomb Ansatz \cite{SS14}.
[See also the main panel of Fig.2(c).]

Thus, we conclude that the perfect Abelian dominance of the string tension, 
i.e., $\sigma_{\rm Abel} \simeq \sigma$, is observed 
on the fine and large-volume lattice of $32^4$ at $\beta =6.4$.

\vspace{0.4cm}

\begin{table}[h]
\caption{
Fit analysis for $V(r)-V_{\rm Abel}(r)$ on the $32^4$ lattice at $\beta =6.4$. 
We list the best-fit parameter set $(\sigma, A, C)$ 
in the Coulomb-plus-linear Ansatz, and the best-fit parameter set 
$(A, C)$ in the Coulomb Ansatz with $\sigma=0$.
These fit results indicate that $V - V_{\rm Abel}$ 
takes a pure Coulomb form and has zero string tension.
}
\begin{tabular}{ccccc}
\hline
 &  $\sigma$ $[a^{-2}]$  & $A$ & $C$ $ [a^{-1}]$ & $\chi^2/N_{\rm df}$ \\
\hline
$V - V_{\rm Abel}$ & $-$0.00017(22)  & 0.223(3) & 0.433(2) & 0.86\\
$V - V_{\rm Abel}$ &  --- & 0.221(2) & 0.432(1) & 0.84\\
\hline
\end{tabular}
\end{table}

\subsection{Simple nontrivial summation formula}

In contrast to Abelian dominance for the long-distance confinement properties, 
there is a significant difference between $V(r)$ and $V_{\rm Abel}(r)$ 
at short distances.
For the accurate analysis of the short-distance behavior of the potential, 
the fine lattice at $\beta$=6.4 is preferable.

As shown in Fig.2(b), 
we newly find a simple but nontrivial summation formula \cite{SS14} of
\begin{equation}
V(r) \simeq V_{\rm Abel}(r) + V_{\rm off}(r)
\label{eq:sum}
\end{equation}
among the Q$\bar {\rm Q}$ potential $V(r)$, 
the Abelian part $V_{\rm Abel}(r)$, 
and the off-diagonal part $V_{\rm off}(r)$.

This summation formula~(\ref{eq:sum}) indicates that 
the significant difference between $V(r)$ and $V_{\rm Abel}(r)$ 
at short distances 
is almost complemented by the off-diagonal part $V_{\rm off}(r)$.
In the non-Abelian theory, however, 
this simple summation formula ~(\ref{eq:sum}) is fairly nontrivial, 
because the link-variables are not commutable, e.g., 
\begin{eqnarray}
\prod_C M_\mu (s) u_\mu (s) \ne 
\prod_C M_\mu (s) \prod_C u_\mu (s),
\end{eqnarray}
and this non-Abelian nature generally leads to 
$W_C \left[ U_\mu(s) \right] 
\ne W_C \left[ M_\mu(s) \right] \cdot W_C \left[ u_\mu(s) \right]$.
Nevertheless, the summation formula ~(\ref{eq:sum}) is observed 
especially on the fine and large-volume lattice of $32^4$ at $\beta=6.4$.

\section{Summary and concluding remarks}

We have studied MA projection of quark confinement in SU(3) QCD 
on a fine and large-volume lattice of $32^4$ at $\beta$=6.4. 
We have investigated the static Q$\bar{\rm Q}$ potential $V(r)$, 
its Abelian part $V_{\rm Abel}(r)$, and its off-diagonal part 
$V_{\rm off}(r)$, from the on-axis lattice data. 
Remarkably, we have found almost perfect Abelian dominance 
of quark confining force, i.e., 
$\sigma_{\rm Abel} \simeq \sigma$ for the string tension, 
on the fine and large-volume lattice.
In addition, we have newly found a simple nontrivial summation relation of 
$V(r) \simeq V_{\rm Abel}(r) + V_{\rm off}(r)$. 

Thus, in spite of the non-Abelian nature of QCD, 
quark confinement is entirely kept in the Abelian sector of QCD 
in the MA gauge.
In other words, Abelianization of QCD can be realized 
without loss of quark confining force by MA projection, 
which can partially reduce the difficulty 
stemming from non-Abelian nature of QCD.
In any case, such infrared Abelianization scheme of QCD 
would be meaningful to understand the quark confinement mechanism 
in the non-Abelian gauge theory of QCD.

\begin{theacknowledgments}
H.S. thanks Prof. V.~G.~Bornyakov for his valuable suggestions.
H.S. is supported by the Grant for Scientific Research [(C) No.23540306] 
from the Ministry of Education, Science and Technology of Japan.
N.S. is supported by a Grant-in-Aid for JSPS Fellows (Grant No.~250588)
and by RIKEN iTHES Project.
The lattice QCD calculations were partially performed on NEC-SX8R 
at Osaka University.
\end{theacknowledgments}



\bibliographystyle{aipproc}   

\bibliography{sample}

\IfFileExists{\jobname.bbl}{}
 {\typeout{}
  \typeout{******************************************}
  \typeout{** Please run "bibtex \jobname" to optain}
  \typeout{** the bibliography and then re-run LaTeX}
  \typeout{** twice to fix the references!}
  \typeout{******************************************}
  \typeout{}
 }



\end{document}